Original Paper

# A Surveillance Infrastructure for Malaria Analytics: Provisioning Data Access and Preservation of Interoperability

Mohammad Sadnan Al Manir[1], BSc, MS; Jon Haël Brenas[2], BSc, MS, PhD; Christopher JO Baker[1,3], PhD; Arash Shaban-Nejad[2], BSc, MPH, MSc, PhD

[1]Department of Computer Science, University of New Brunswick, Saint John, NB, Canada
[2]Oak Ridge National Laboratory Center for for Biomedical Informatics, Department of Pediatrics, The University of Tennessee Health Science Center, Memphis, TN, United States
[3]IPSNP Computing Inc, Saint John, NB, Canada

**Corresponding Author:**
Arash Shaban-Nejad, BSc, MPH, MSc, PhD
Oak Ridge National Laboratory Center for for Biomedical Informatics
Department of Pediatrics
The University of Tennessee Health Science Center
50 N Dunlap Street, R492
Memphis, TN, 38103
United States
Phone: 1 901 287 5836
Email: ashabann@uthsc.edu

## Abstract

**Background:**   According to the World Health Organization, malaria surveillance is weakest in countries and regions with the highest malaria burden. A core obstacle is that the data required to perform malaria surveillance are fragmented in multiple data silos distributed across geographic regions. Furthermore, consistent integrated malaria data sources are few, and a low degree of interoperability exists between them. As a result, it is difficult to identify disease trends and to plan for effective interventions.

**Objective:**   We propose the Semantics, Interoperability, and Evolution for Malaria Analytics (SIEMA) platform for use in malaria surveillance based on semantic data federation. Using this approach, it is possible to access distributed data, extend and preserve interoperability between multiple dynamic distributed malaria sources, and facilitate detection of system changes that can interrupt mission-critical global surveillance activities.

**Methods:**   We used Semantic Automated Discovery and Integration (SADI) Semantic Web Services to enable data access and improve interoperability, and the graphical user interface-enabled semantic query engine HYDRA to implement the target queries typical of malaria programs. We implemented a custom algorithm to detect changes to community-developed terminologies, data sources, and services that are core to SIEMA. This algorithm reports to a dashboard. Valet SADI is used to mitigate the impact of changes by rebuilding affected services.

**Results:**   We developed a prototype surveillance and change management platform from a combination of third-party tools, community-developed terminologies, and custom algorithms. We illustrated a methodology and core infrastructure to facilitate interoperable access to distributed data sources using SADI Semantic Web services. This degree of access makes it possible to implement complex queries needed by our user community with minimal technical skill. We implemented a dashboard that reports on terminology changes that can render the services inactive, jeopardizing system interoperability. Using this information, end users can control and reactively rebuild services to preserve interoperability and minimize service downtime.

**Conclusions:**   We introduce a framework suitable for use in malaria surveillance that supports the creation of flexible surveillance queries across distributed data resources. The platform provides interoperable access to target data sources, is domain agnostic, and with updates to core terminological resources is readily transferable to other surveillance activities. A dashboard enables users to review changes to the infrastructure and invoke system updates. The platform significantly extends the range of functionalities offered by malaria information systems, beyond the state-of-the-art.

*(JMIR Public Health Surveill 2018;4(2):e10218)*   doi: 10.2196/10218

## Introduction

### Malaria Surveillance Today

Malaria is an infectious disease with significant impact on developing countries. In 2016 alone, it caused 445,000 deaths worldwide, and globally around 216 million cases of malaria have been reported in 91 countries [1]. Populations in sub–Saharan African countries are most susceptible, with 80% of observed cases and recorded deaths worldwide [2,3]. Whereas a decrease in case incidence has been observed [1] since 2010, the rate of decline appears to have stalled, in part due to lack of adequate surveillance and intervention programs. An essential prerequisite for accelerating the decline of the disease and optimally targeting resources is an efficient surveillance infrastructure that can reliably deliver robust datasets. Poor data quality and sparseness and the coordination between different surveillance systems are ongoing challenges for the malaria surveillance community [4]. The increasing number of stakeholders, including international organizations, governments, nongovernmental organizations, and private sectors [5], that contribute to gathering the data can lead to siloed heterogeneous information systems and data sources that need to be integrated [6]. Overall, infrastructures for malaria surveillance are brittle [7], and stakeholders do not have sufficient confidence in the aggregated datasets to adopt any conclusions that could be meaningfully derived [8].

A comprehensive study [9] showed that malaria information systems that can collect, store, and analyze data and provide feedback to developers based on real-time information are few. Moreover, the most effective systems are still limited by the absence of real-time data aggregation, inconsistent decision support, and low levels of resolution, such as no mapping to households and no extrapolation across geographic borders. In specific cases where an information system [10-12] has been upgraded with improved visualization and reporting tools, other challenges remain unresolved. Data entered from field stations are managed centrally, and updates occurring at the central data source are not reflected in the field level immediately. Also, the options to generate reports are generally predetermined; hence, ad hoc queries cannot be answered without significantly overhauling the systems. Consequently, the types of surveillance queries that can be run to derive actionable knowledge in a timely manner are relatively few.

Furthermore, 11 widely used Web platforms were studied to assess how internet and Web technologies are used in the fight against malaria [13]. The elements of this study were focused on data, metadata, Web services, and categories of users. The results revealed that, although heterogeneous spatiotemporal malaria data came from multiple disciplines, they were rarely updated dynamically, no metadata were used to standardize them, Web services were inflexible for reuse and nonstandardized, and the platforms primarily served the scientific communities. The authors identified that, to improve these systems, interoperability through standardization is necessary. The World Health Organization's global technical strategy [14] also identified that surveillance challenges in sub–Saharan African countries exist partly because mechanisms for facilitating data integration from distributed data silos and ensuring system interoperability are lacking.

### Beyond the State-of-the-Art

Indeed, the malaria surveillance community is not alone in facing these challenges, and many other communities are investigating how to bring distributed datasets together in real time to support decision making. Researchers in other domains have sought to introduce guidelines for ensuring that source data are published in ways that ensure they are findable, accessible, interoperable, and reusable [15], albeit no specific technical solutions are proposed or mandated. In addition, standards and software prototypes have emerged specifically from the Semantic Web community [16] but, given the nature of these challenges, it is the combined benefits of applying guidelines and technical solutions that would provide a transformational development for surveillance practitioners.

The specific challenge of interoperability has two dimensions, namely structural and semantic interoperability. Structural or syntactic interoperability can be achieved by defining common syntax and formats for data exchange. For example, if two systems such as the Malaria Atlas Project [17] and Africa Development Indicators [18] are using demographic data from a census stored in another database, having interoperability between the two systems means that users of these databases can mutually access and reuse census data without having to store them locally or reformat the data on import. Semantic interoperability, on the other hand, is much harder to achieve. The goal is to ensure that the integrity and meaning of the data is preserved throughout the integration process. This can be achieved by mapping data to standardized vocabularies or terms in ontologies [19], such as the Vector Surveillance and Management Ontology (VSMO) [20], Infectious Disease Ontology-Malaria (IDOMAL) [21], or Mosquito Insecticide Resistance Ontology (MIRO) [22], which provide means to formally model a domain of interest using concepts in the domain, relations among the instances of concepts, and complex logical axioms. Two sets of data mapped to the same ontology terms or vocabularies can be regarded as having the same meanings. Ontologies facilitate semantic interoperability in integrated biomedical and health systems and can also be used to make malaria data sources comparable. Semantic interoperability achieved through mapping to community standard ontological terms is therefore an essential property for surveillance routines and enables more integrated access to data.

Overcoming the challenge of distributed data access has relied on established technologies such as Web services, but interoperability is still lacking in many implementations. In recent work, Web service-based data access and interoperability challenges have been tackled together using Semantic Web service infrastructures [23]. The benefits of this approach include enhanced findability of distributed online resources, easy construction of workflows to fetch data from multiple Web services, and access to source data in interoperable formats. Existing deployments of this approach [23,24] as part of the Semantics, Interoperability, and Evolution for Malaria Analytics (SIEMA) [23] platform have aimed to address the interoperability between multiple dynamic and evolving malaria data sources. Early trials have reported advanced query options [24] for end users without advanced technical skills. In addition, SIEMA seeks to manage infrastructure changes, where *change management* is defined as the process of preserving the integrity and consistency of an integrated system while keeping the information up-to-date. Here the primary activities during change management are detection, representation, validation, traceability, and rollback, as well as reproduction of the changes [25]. SIEMA has used authentic scenarios (use case) provided by the Uganda Ministry of Health to illustrate effectiveness in providing seamless access to distributed data and preservation of interoperability between online resources.

### Objective

We introduce a prototype surveillance and change management platform, known as SIEMA, built from a combination of third-party tools, community-developed terminologies, and custom algorithms. We illustrate the methodology and core infrastructure used to facilitate interoperable access to distributed data sources using Semantic Automated Discovery and Integration (SADI) [26] Semantic Web services. We show a dashboard that reports on terminology changes that can render SADI services inactive, jeopardizing system interoperability, allowing end users to control and reactively rebuild services to preserve interoperability and minimize service downtime.

## *Methods*

The SIEMA surveillance platform relies on the coordination and customization of a number of existing frameworks, and software and custom-developed algorithms. The architecture diagram in Figure 1 describes these resources in an abstract representation of the SIEMA platform first introduced by Brenas et al [23]. We briefly describe the key resources below.

### Semantic Automated Discovery and Integration Semantic Web Service

SADI is a representational state transfer (RESTful) Web service framework that provides a set of conventions for creating Semantic Web services. The framework uses resource description framework schema (RDF[S]) [27] and Web Ontology Language (OWL) [28] standards for data representation and modeling, and HTTP-based recommendations (GET, POST) for interacting with the services. SADI services consume and produce RDF instances of OWL classes where the input data are decorated until they become an instance of the output OWL class. The services are deployed in a registry and can be automatically discovered, orchestrated, and invoked to return data in RDF on the query clients SHARE [29] and HYDRA [30,31].

### Source Data and Standard Terminologies

Our research focuses on middleware for enabling discovery of datasets and tools for agile query composition, but deployments of these methodologies beyond prototypes will require full access to malaria data. Existing data repositories such as the Scalable Data Integration for Disease Surveillance [32] or Global Malaria Mapper [33] would be target resources for building service descriptions and including them in a generic registry. However, a recent study [13] suggested that in order to facilitate change management and semantic interoperability these data sources must be represented by standard terminologies related to their domain. Rather than defining new terminologies, it is a best practice to use community-adopted domain terminologies, if available, for defining the services. For this purpose, several ontologies are leveraged, including VSMO [20], IDOMAL [21], MIRO [22], and the Public Health Ontology [34]. The Clinical Data Interchange Standards Consortium [35] also offers ontologies.

### Query Clients for Semantic Automated Discovery and Integration Services

SHARE is a specialized open source query client that enables end users to discover, plan, and orchestrate SADI services in a registry and invoke them automatically from SPARQL [36] queries. The services are discovered by finding a match to the predicates attached to the input URIs after comparing both the semantic descriptions of the services' input and the output OWL classes. Each triple pattern in the SPARQL query is resolved by checking the predicates in the service registry. The order of invocation of services is strategized during the planning stage. The services are then orchestrated and invoked to execute the desired tasks. The output triples are produced through the binding of subject-object values to the corresponding variables in the SPARQL query. To generate the complete output, every triple in the query is resolved. HYDRA is a commercial query client for SADI services, developed by IPSNP Computing Inc. HYDRA can be used as a Java application programming interface, as a simple command line application, or through an intelligent graphical user interface (GUI) supporting ad hoc query composition by nontechnical users by combining Google-style keyword-based querying with query graph editing. Instead of queries being written in SPARQL, the GUI allows using a graph representation that is easier to understand.

### Agent-Based Analytics Dashboard

Detecting changes in the source data schema, as well as the domain and service ontologies, is a prerequisite step for change management. Studies on the evolution of large domain ontologies [37] have shown that the vocabularies are subject to change [38] because they have shared authorship, and they evolve to represent new knowledge. Because of the high degree of dependency between system components in a relational data schema, even a small change in a data schema could result in cascading impacts across an application stack. In previous work, we presented features of a Web-based analytics dashboard for detecting and reporting changes [39].

**Figure 1.** Architecture of the Semantics, Interoperability, and Evolution for Malaria Analytics (SIEMA) surveillance framework. GUI: graphical user interface; I/O: input/output; RDF: resource description framework; SADI: Semantic Automated Discovery and Integration.

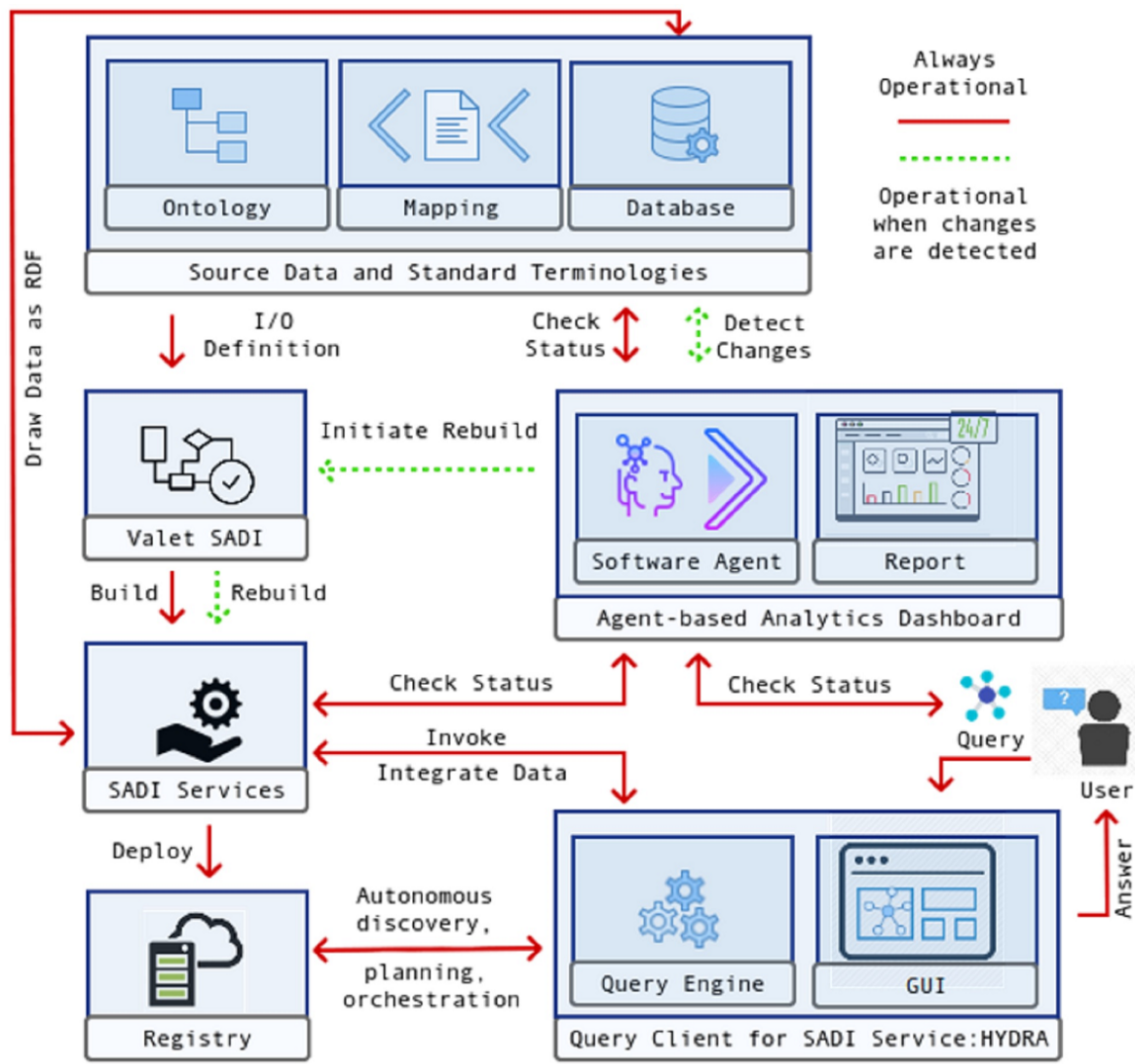

Underpinning the dashboard, software agents enable two key actions: (1) detecting changes and identifying their types, and (2) restoring the modified component to an operational state through repair and rebuilding. The types of changes we accommodate are addition (ie, extension), deletion (ie, obsoleting), and renaming (ie, refining) of components in domain ontologies and service ontologies [40,41], such as concepts, property restrictions, individuals, and axioms. Changes in the source data, such as tables and their attributes, data types, and indices, can also be reported [42]. Likewise, changes to saved queries and the corresponding service dependencies can be detected. Brenas et al [23] report further details of the agents designed to capture changes in various components of the infrastructure and their interaction with other software agents in the system.

**Valet SADI**

Valet SADI [43] generates SADI services automatically based on the formal definition of a service and rules for mapping a data source to ontology terms. The knowledge required to interpret the data is expressed using rules scripted in positional-slotted object-applicative (PSOA) RuleML [44]. These rules define how each element coming from a data source has to be matched to the vocabulary defined in the ontologies. Valet SADI, with the help of the domain ontologies and the semantic mapping rules, rewrites the declarative input and output definitions of a service into a query that is specific to the source data schema, such as Structured Query Language (SQL), and creates the complete Java code implementing the service functionality. It facilitates the generation and deployment of a service in a matter of seconds, which is essential in the context of change management, where uptime is a priority.

## Results

**Service-Based Querying of Malaria Data**

The implementation presented in this paper is inspired by the objectives stated in the National Malaria Control Program by the Ugandan Ministry of Health [45]. The questions, shown in Textbox 1, based on these objectives resemble authentic queries for a surveillance practitioner in the malaria domain. In this research, we converted these questions into formal queries by semantically presenting them as one or more SADI services. This section illustrates how the queries can be represented using a combination of SADI services on a graphical query canvas of the HYDRA query engine.

### *Questions for Malaria Surveillance*

We selected primarily a series of widely used malaria control interventions. The importance of indoor residual spraying has been well established by numerous studies throughout the world, especially in Africa [45,46], as an effective way of killing the mosquitoes that transmit malaria. Hence, the first question in Textbox 1 is about indoor residual spraying with a specific insecticide. We use this question as a running example in the rest of this paper. The use of insecticide-treated bed nets, alongside indoor residual spraying, is the second most effective intervention against malaria. Q2 in Textbox 1 is more complex to formalize and answer because it requires aggregating information about bed nets from geographic locations, insecticides used, and a comparison of the status of mosquito populations within a certain time frame in a specific location. Q3 in Textbox 1 anticipates that a malaria surveillance practitioner will be interested in predicting future trends and outbreaks to improve the allocation of resources where they are most needed. We consider these three queries as authentic target queries and adopted them as a basis for illustrating and evaluating the capabilities and appropriateness of the proposed technical approach.

### *Building Semantic Automated Discovery and Integration Services*

To answers the questions, we created and deployed a list of SADI services in a registry. We focus below on how to create those services.

#### **Source Data and Vocabulary**

Vocabularies from one or more domain ontologies are used to define the input and the output of a service. The data schema of the source data is also necessary. The vocabularies and the data schema in Figure 2 are used to define the SADI services. In the domain ontology, the hierarchy of classes includes VSMO:0001957 and MIRO:10000239, each of which represents the concept of spraying an insecticide. The object properties *has_insecticide* and *located_in* link the records of spraying to the records of insecticides and geographic regions. The data property *has_name* represents the literal values of names of the specific spraying activity, the geographic region, and the insecticide used. In the data schema, the table labeled spraying contains information about all spraying activities; table geographicregion contains information about the locations where the insecticides were sprayed, and table insecticide contains insecticide information. In table spraying, the integer-valued attribute *id* is used to identify each spraying activity, attribute *name* represents the name of the activity, *location.id* refers to the location where the activity was performed, year represents the time it was performed, and *insecticide.id* refers to the insecticide used in the spray. In table geographicregion, the integer-valued attribute *id* represents each region and the attribute *name* represents the name of that region. Finally, in table insecticide, the integer-valued attribute *id* represents a pesticide, the attribute *name* identifies the name of the pesticide, and the attribute *mode.of.action* decides whether the pesticide is effective on mosquitoes that come in *contact* with it or both *contact & airborne*. The attributes *location.id* and *insecticide.id* act as foreign keys to the region and insecticide, respectively.

#### **Description of Semantic Automated Discovery and Integration Services**

The names of SADI services are expressed in two different forms: (1) allX, which retrieves all information regarding X without expecting any input, and (2) getYByZ, which retrieves Y based on the input Z. The input and the output of every service are defined in a service ontology using the terminologies from the domain ontologies. One such service is getInsecticideIdByIndoorResidualSprayingId, which takes an instance of spraying as input, which is any element whose *type* is indoor residual spraying. The service returns an *id* representing an insecticide. The input is decorated explicitly by the relation *has_insecticide*. Figure 3 defines the descriptions of this service.

Another service is getNameByInsecticideId, which takes an instance of an insecticide as input. The service returns a string as output, representing the name of the insecticide in the data, decorating the input by the relation *has_name*. Figure 3 shows the input-output descriptions of this service.

**Textbox 1.** Questions inspired by the National Malaria Control Program by the Ugandan Ministry of Health.

> Q1. Which indoor residual sprayings used permethrin as an insecticide?
>
> Q2. Which districts of Uganda that used permethrin-based long-lasting insecticide-treated nets in 2015 saw a decrease in *Anopheles gambiae* s.s. population but no decrease of new malaria cases between 2015 and 2016?
>
> Q3. What are the future high-risk areas and at-risk time periods in Uganda?

**Figure 2.** Snapshot of source data schema and domain ontologies.

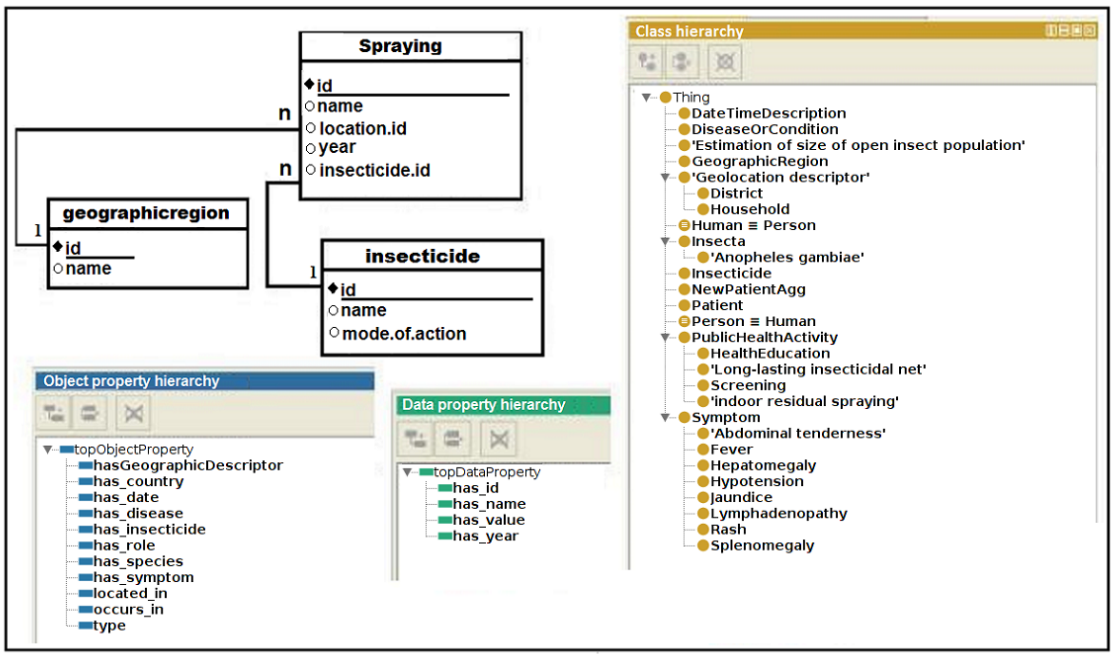

**Figure 3.** Input (left) and output (right) descriptions of the services getInsecticideIdByIndoorResidualSprayingId (top) and getNameByInsecticideId (bottom).

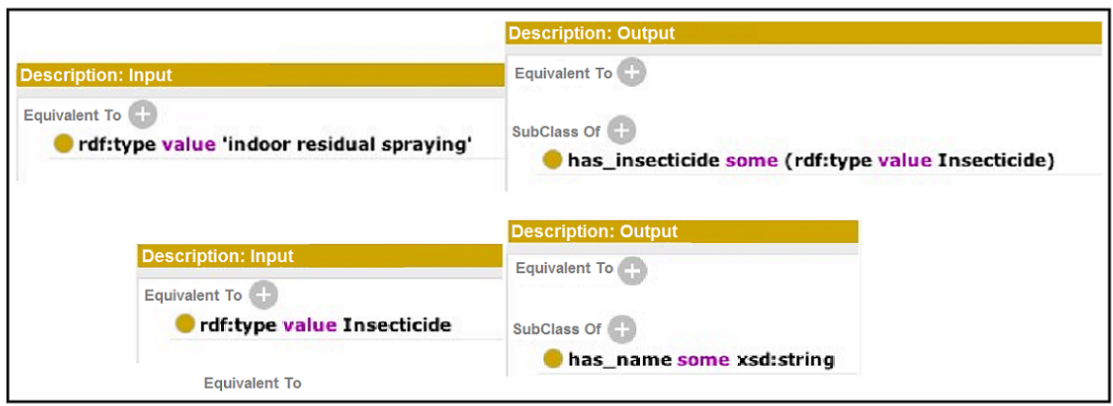

**Figure 4.** A fragment of the registry of Semantic Automated Discovery and Integration (SADI) services.

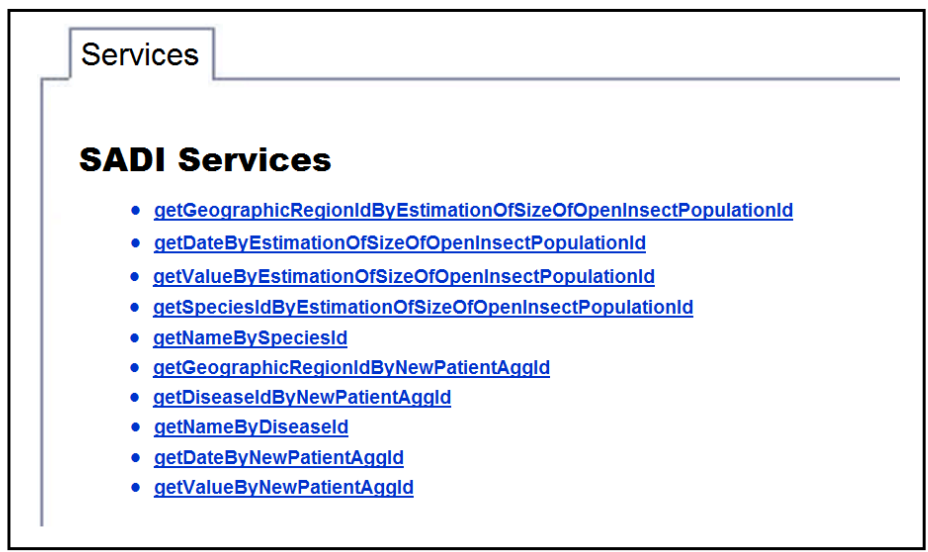

**Textbox 2.** Example of mapping rules in positional-slotted object-applicative RuleML.

```
1  Forall ?insecticideID (identityForInsecticideToinsecticideID(
2           identityForInsecticide(?insecticideID)) = ?insecticideID)
3  Forall ?P (identityForInsecticide(identityForInsecticideToinsecticideID(?P))= ?P)
4  Forall ?id ?name ?mode.of.action (
5    Insecticide(identityForInsecticide(?id)) :-
6     db_insecticide(?id ?name))
7  Forall ?id ?name ?mode.of.action (
8    has_name(identityForInsecticide(?id) ?name) :-
9     db_insecticide(?id ?name))
```

### Service Registry

Once the services have been generated, they are stored in a service registry. Figure 4 shows a fragment of the service registry that we use for malaria surveillance. HYDRA uses the registry to discover the services that will be called during the execution of the queries.

### Provisioning of Semantic Automated Discovery and Integration Services

Specification and building of SADI services can be cumbersome, error prone, and tedious for nontechnical end users. Full implementation details are outside the scope of this paper; however, we encourage readers to look at the details in Brenas et al [23]. Here, we used Valet SADI, which accelerates the building of SADI services, to automatically implement the services from their input and output descriptions. Riazanov et al [30] demonstrated how rules expressed in PSOA RuleML semantically map the underlying source data schema to the vocabularies of the domain. As source data are usually not expressed in the language defined by the ontologies, such rules have been used to explain how to interpret the data. Textbox 2 shows an example of such mapping rules. Once these rules are available, Valet SADI uses them to access the data by rewriting the description of the services into correct queries for the mapped data schema.

### Building Queries

To illustrate query building, consider Q1 in Textbox 1, which is easily translated to a query. Figure 5 shows its graph representation. The GUI of HYDRA makes it easier to build queries because relationships are easier to understand and process in graph form than in SPARQL syntax.

Services in (3) and (4) are described above, while the service in (1) retrieves all identifiers of public health activities in Uganda, and the service in (2) retrieves the names of these activities. The branch on the right in Figure 5 uses the services (1), (3), and (4), while the left branch requires services (1) and (2). Queries are constructed in an incremental fashion. The root node starts with the service allPublicHealthActivities, which can then be extended either to the left or to the right. The service getNameByPublicHealthActivityId is used on the left to name the intervention represented by the variable *inter_name*. The root node is then decorated by the property *has_insecticide* with the description of the service getInsecticideIdByIndoorResidualSprayingId to represent an identifier of an insecticide. Finally, the service getNameByInsecticideId represents the name of the insecticide with a value *Permethrin*. As queries become more complex, the advantage of graph form over the raw SPARQL syntax becomes evident. Figure 6 shows the graph form of Q2. Although the query graph is significantly more complex and larger than the query in Q1, it is much easier to understand for any user than the raw SPARQL representation. Due to space constraints, we do not describe each service used in the query graph of Q2 or of Q3.

The query in Figure 5 calls on four distinct SADI services:

1. allPublicHealthActivities,
2. getNameByPublicHealthActivityId,
3. getInsecticideIdByIndoorResidualSprayingId, and
4. getNameByInsecticideId.

**Figure 5.** Graph representation of a query for the question "Which indoor residual sprayings used permethrin as an insecticide?" prepared on the HYDRA graphical user interface.

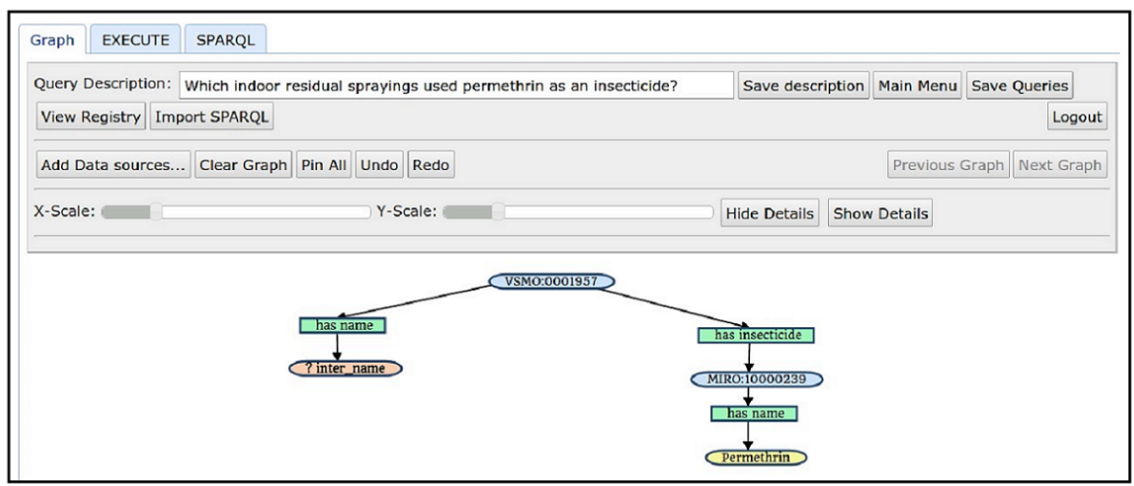

**Figure 6.** Graph representation of a query for the question "Which districts of Uganda that used permethrin-based long-lasting insecticide-treated nets in 2015 saw a decrease in *Anopheles gambiae* s.s. population but no decrease of new malaria cases between 2015 and 2016?".

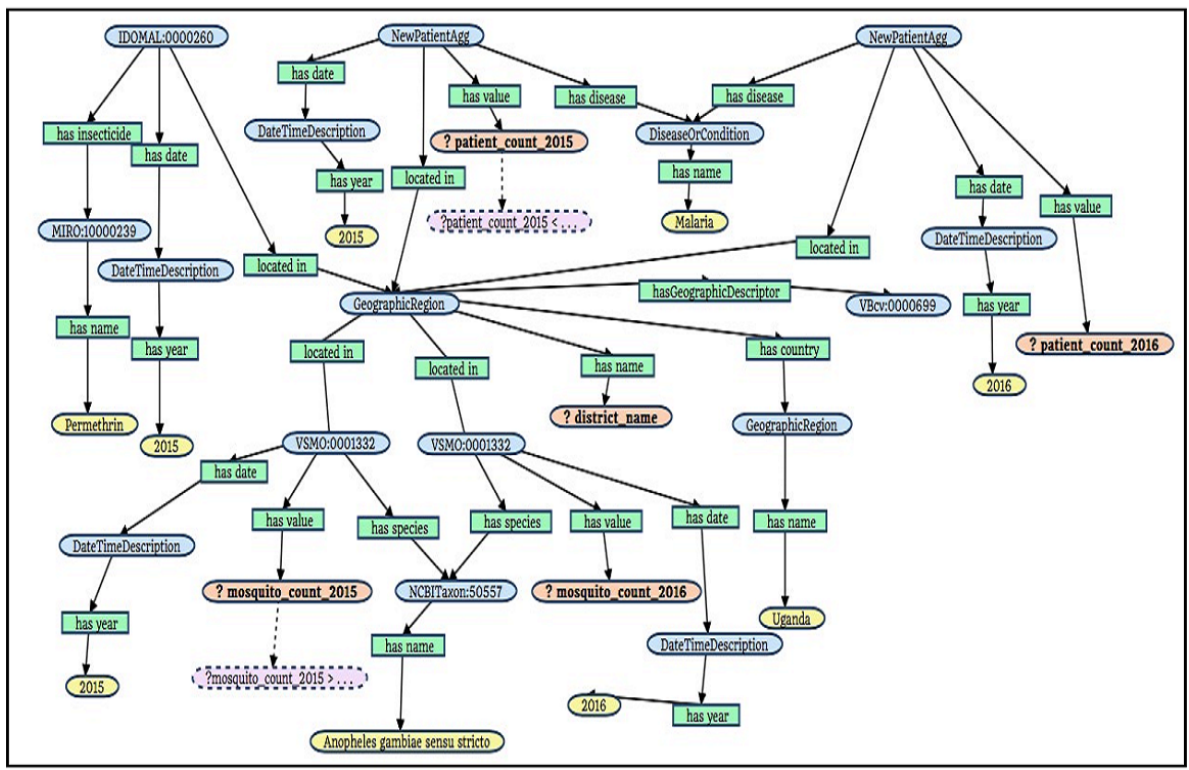

**Figure 7.** Graph representation of the query "Which indoor residual spraying used permethrin as an insecticide and which kind of mosquitoes will be affected by it?".

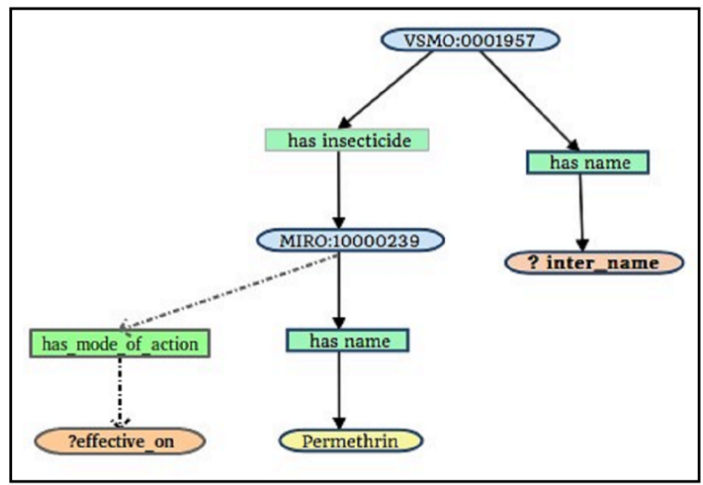

### Change Management

The previous section outlined how, using Semantic Web services, it is possible to answer complex questions relevant to malaria surveillance. Special attention is needed before considering the introduction of a new methodology in a dynamic context where data and middleware are not static. Several possible changes that could occur in a malaria surveillance framework have been described and classified according to the degree to which they affect data access and their likelihood to affect interoperability of the system [23]. In the following, we illustrate an approach to address the notion of change management through the introduction of change detection tools and triggers for the reactive rebuilding of service to ensure service uptime. Specifically, we discuss the addition of terms that are used in service ontologies and the use of a reporting tool for detecting and displaying these types of changes.

#### Changes in the Service Ontologies

Whenever a definition of an existing service is modified, the associated service ontology is changed, but the code implementing the service remains unchanged. As a result, when a service is invoked during the execution of a query, it does not return the anticipated output because the terms used in the code are incompatible with the new definition in the service ontology. In the SIEMA framework, the change capture agent implemented within the dashboard detects the changes in the terms used in the service ontology by comparing the modified version of the ontology with the one it was modified from. The role of the change capture agent can be illustrated in the case of a term addition.

To illustrate a scenario involving the addition of new terms, consider the query in Figure 5 again. The mode of action, mentioned in the World Health Organization's recommendation [47], plays an important role in choosing insecticides for indoor residual spraying against malaria vectors.

Let us assume that the end user is interested in the query "Which indoor residual spraying used permethrin as an insecticide and which kind of mosquitoes will be affected by it?" Figure 7 shows the graph representation of such a query. To be able to answer this query, the existing service getNameByInsecticideId can be modified by adding the term *has_mode_of_action* and the datatype xsd:string that it returns. The new output description in Figure 8 is defined by adding the two terms to the old description.

A change capture agent detects the changes of two terms, identifies them as addition, and displays them on the dashboard in a tabular form. Table 1 shows the consolidated information about the changes on the dashboard, which displays the time the changes were detected in the *Timestamp* column; the textual description of the change in the *Description of change* column; the types of changes identified as either addition in the *Entity added* column, deletion in the *Entity deleted* column, or new name in the *Entity renamed* column; the list of services affected in the *Affected service* column; and the list of queries affected in the *Affected query* column.

#### Status of Services

At any time, the status of all deployed services in the registry is displayed on the dashboard. Services can be either active, which can be used in queries, or inactive, which need to be repaired before using them again. Figure 9 displays a list of active and inactive services in green and red, respectively. The URI of the service is tabulated in the *Service URI* column, the description is under *Description*, the time the service was created is under *Time of creation*, and the time the service was rebuilt is under *Time of rebuild*. The final column is *Request rebuild*, which allows for placing an inactive service in a queue to be rebuilt and subsequently redeployed in the service registry. The getNameByInsecticideId service is shown in red because it became inactive due to the addition of terms in its output definition.

#### Reacting to Changes

The addition of terms to the definition of an active service renders the service description incompatible with the target functionality and existing service code, and renders associated queries dysfunctional. To resolve the inconsistency, it is necessary to repair and rebuild the services in line with the new requirements. Specifically, the end user now wants to access the data that were not previously available from a service, namely, in this example, the mode of action of an insecticide. It is thus necessary to reimplement the service corresponding to the altered service description ensuring that the domain ontologies, the data schemata, and the PSOA semantic mapping rules underpinning the service are accurate and will support the new target functionality.

Given that a data resource contains the information about the mode of action of insecticides, the key question is whether the semantic mapping rules already map those data to an existing concept or relation of the domain ontologies. If that is the case, then all components required to rebuild the services exist, and it is possible to proceed to the next step in the Valet SADI rebuild. Otherwise, it is necessary to identify missing rules and add them, or extend a local domain ontology with a missing concept or relation that exists in the service ontology. Once this is done, a rule must be created to define a new mapping and to make rebuilding the service possible.

### *Rebuilding the Services Using Valet SADI*

By leveraging Valet SADI's autogeneration capability, the damaged service can be quickly rebuilt and deployed once changes are detected and identified, and a rebuild is requested. To illustrate this, we refer to the query shown in Figure 7, which calls services retrieving data from the source database schema and list the changes below.

**Domain Ontology**

In the domain ontology, the data property *has_mode_of_action* is added. As a result, the service ontology becomes compatible and supports the extended query, as it only uses existing concepts and properties.

**Positional-Slotted Object-Applicative Semantic Mapping Rules**

The PSOA rules are also modified to populate the newly added data property *has_mode_of_action* as Textbox 3 shows. The lines 22'-24' are added in contrast to the original version in Figure 5. Valet SADI can rewrite the modified input and output descriptions with the help of the mapping rules, and generate the SQL query and the complete program code automatically to fetch answers from the source data. The services are thus rebuilt and redeployed in the service registry so that the modified query returns the correct answers. Figure 10 shows the updated status of the services. The service getNameByInsecticideId is now shown in green because the service was rebuilt by Valet SADI and redeployed in the service registry. The exact time the service was rebuilt is tabulated in the *Time of rebuild* column.

**Figure 8.** Old (left) and new (right) output description of the service getNameByInsecticideId.

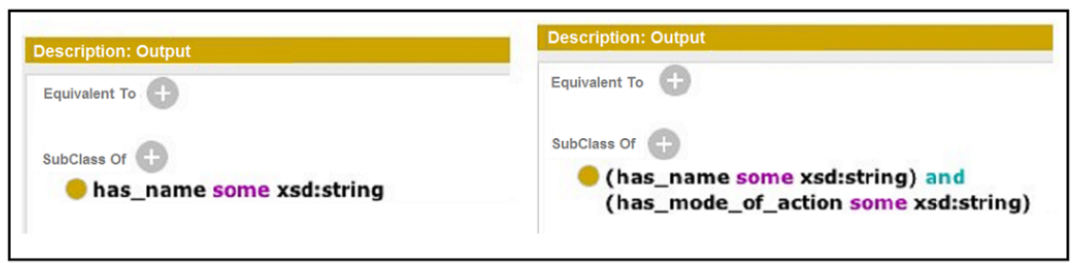

**Table 1.** Detection and identification of changes in the service ontology.

| Timestamp | Description of change | Entity added | Entity deleted | Entity renamed | Affected service | Affected query |
| --- | --- | --- | --- | --- | --- | --- |
| 2018-01-21T14:33:08 | An entity is added to the output definition | has_mode_of_action | N/A[a] | N/A | getNameByInsecticideId | Which indoor residual sprayings used permethrin as an insecticide? |
| 2018-01-21T4:33:08 | An entity is added to the output definition | xsd:string | N/A | N/A | getNameByInsecticideId | Which indoor residual sprayings used permethrin as an insecticide? |

[a]N/A: not applicable.

**Figure 9.** Status of Semantic Automated Discovery and Integration (SADI) services. Active services are shown in green and inactive services in red.

**Textbox 3.** Fragment of the updated positional-slotted object-applicative file.

12' Group (

13'  Forall ?insecticideID (identityForInsecticideToinsecticideID(

14'           identityForInsecticide(?insecticideID)) = ?insecticideID)

15'  Forall ?P (identityForInsecticide(identityForInsecticideToinsecticideID(?P)) = ?P)

16'  Forall ?id ?name ?mode.of.action (

17'   Insecticide(identityForInsecticide(?id)) :-

18'    db_insecticide(?id ?name ?mode.of.action))

19'  Forall ?id ?name ?mode.of.action (

20'   has_name(identityForInsecticide(?id) ?name) :-

21'    db_insecticide(?id ?name ?mode.of.action))

22'  Forall ?id ?name ?mode.of.action (

23'   has_mode_of_action(identityForInsecticide(?id) ?mode.of.action) :-

24'    db_insecticide(?id ?name ?mode.of.action))

25' )

**Figure 10.** Status of services after being rebuilt by Valet SADI.

## Discussion

Surveillance remains a challenge for the malaria community, and many factors play a role in limiting access to relevant data resources for analysis and reuse [13]. We investigated the suitability of a solution to the challenges of system interoperability, distributed data access, semantic integration of data, and semantic support for query composition.

**Semantics, Interoperability, and Evolution for Malaria Analytics**

The SIEMA framework comprises several technologies and standards and is further customized to address the proposed targeted needs and interests of surveillance practitioners. The contribution takes three main directions. First, using SADI Web services allows for easy access to distributed data. This task is simplified further using Valet SADI, which enables a programmer to create services in an efficient and straightforward way. Second, due to the user interface features of the HYDRA query engine, SIEMA offers end users a more appealing way to build surveillance queries. HYDRA's ability to discover and call the services that are needed for a query permits the user to simply use the data as an abstract construct without having to look at its actual structure. Third, to make the system more robust and flexible, a dashboard has been introduced. The dashboard informs users when changes have occurred that render the services or queries inactive. This enables users to know which queries may no longer be reliable and to identify which parts of the service infrastructure must be rebuilt to restore it to its fully interoperable state. Deployed together, this combination of technologies offered by the SIEMA framework exhibits key functionalities that are of great value to the community.

Our initial studies in malaria surveillance [23,24] and other domains [30,48,49] showed us that this approach is a viable solution for enhancing interoperability. The successful extension of this methodology to the malaria use case appears promising based on the results from the initial implementation. The addition of a dashboard and Valet SADI extends the capabilities of the SADI framework to make it suitable for change detection, service restoration, and preservation of interoperability.

**Evaluation**

Whereas a systematic evaluation of the many components of this framework, individually and together, is beyond the scope of this initial study, we are aware that other malaria surveillance systems in sub–Saharan African countries have been reported and evaluated in part [12,50] according to attributes recommended by US Centers for Disease Control and Prevention [51]. Indeed, it would be valuable to make some direct system comparisons. Regrettably, few technical details of these systems have been reported, and their key features—such as their architecture, supported data types and data quality, use of data representation standards, capacity for change management, and degree of interoperability supported—have not been disclosed. Without such information, it is not possible to compare them with SIEMA. It is apparent, however, that the systems were designed for centralized data warehousing rather than seamless access to distributed data, and this is a significant distinction.

A brief assessment of SIEMA according to attributes recommended by Centers for Disease Control and Prevention [52], namely simplicity, data quality, flexibility, stability, and timeliness, can be made based on our initial experiments. To address the attribute of *simplicity*, the activity of creating surveillance queries can be assessed. Whereas writing syntactic queries (in SPARQL) for submission to a query client manually requires that a user have expertise with the query language, the HYDRA GUI can save users time and allow more complex surveillance queries to be composed by persons with less technical skill. Figure 5 shows a graph representation of a query built with the GUI, which is in turn translated to SPARQL automatically. To address the attribute of *data quality*, SIEMA's adoption of SADI ensures that the World Wide Web Consortium standard OWL and RDF[S] are used to describe and format every piece of data accessed through the framework. The attribute of *flexibility* is also met, since access to distributed data can be provisioned by using SADI services that are easily deployed in a registry, making them discoverable and readily usable for a variety of ad hoc queries. Likewise, the attributes of *stability* and *timeliness* of the system are met as the implemented dashboard tracks service uptime and reports failures immediately after detecting the faults. System rebuilds with Valet SADI ensure stability so that data can continue to be provided and the system stays operational if changes occur. In this way, interoperability is preserved and any service downtime is kept to a minimum.

Overall, we anticipate that the ongoing trials with the SIEMA framework will give the research and development team further insight into real-world requirements for interoperability and change management in malaria surveillance, leading to further improvements in adaptability and performance. Given the critical need for timely integration of distributed data from multiple heterogeneous sources in an efficient way, we hope to build cooperative partnerships between multiple disciplines, organizations, and sectors. In addition, insights gained from this research are likely transferable to a range of global surveillance projects.

**Conclusion**

We have demonstrated that authentic questions asked in malaria surveillance can be formalized as queries and mapped to a combination of Semantic Web services designed to deliver target data from distributed data sources. We have shown that using SIEMA and leveraging terminologies from community-developed ontologies offer flexibility both for integrating data and for easily composing queries. The developed infrastructure also offers a solution to the problem of change management, an important process for maintaining interoperability and integrity of an integrated surveillance system. Given that changes in the form of addition, renaming, and deletion of terminologies can frequently occur in the face of evolving system requirements, we introduced a change management dashboard. This makes it possible to identify important changes, report on the status of services as a consequence of changes, and offer users the option to rebuild inactive services. The dashboard and service reauthoring routines serve as an important vehicle to maintain system interoperability of mission-critical global surveillance programs. The infrastructure has been implemented and its relevance has been demonstrated with an authentic use case, with the goal of soliciting further requirements from the malaria analytics community. In future work, we will deploy SIEMA on live dynamic data sources.

### Acknowledgments

This work was funded by the Bill and Melinda Gates Foundation (OPP ID # 1162018). A license for the use of HYDRA was provided by IPSNP Computing Inc.

XSL·FO
RenderX

XSL·FO
RenderX

**Abbreviations**

**GUI:** graphical user interface
**IDOMAL:** Infectious Disease Ontology-Malaria
**MIRO:** Mosquito Insecticide Resistance Ontology
**OWL:** Web Ontology Language
**PSOA:** positional-slotted object-applicative
**RDF[S]:** resource description framework schema
**SADI:** Semantic Automated Discovery and Integration
**SIEMA:** Semantics, Interoperability, and Evolution for Malaria Analytics
**SQL:** Structured Query Language
**VSMO:** Vector Surveillance and Management Ontology